\documentclass[aps,preprint,showkeys,showpacs]{revtex4}
\usepackage{amssymb}
\usepackage{amsmath}
\usepackage{amsfonts}

\newtheorem{theorem}{Theorem}

\newtheorem{example}[theorem]{Example}

\begin{document}

\title{The unique non self-referential $q$-canonical distribution \\
and the physical temperature\\
derived from the maximum entropy principle in Tsallis statistics}
\thanks{The 1st version of this preprint is already appeared in
cond-mat/0502298 on 13/02/2005 GMT. The title of the 1st version is replaced
by new one in the 2nd version.}
\author{Hiroki Suyari}
\email{suyari@ieee.org, suyari@faculty.chiba-u.jp}
\affiliation{Department of Information and Image Sciences, Faculty of Engineering, Chiba
University, 263-8522, Japan}

\begin{abstract}
The maximum entropy principle in Tsallis statistics is reformulated in the
mathematical framework of the $q$-product, which results in the unique non
self-referential $q$-canonical distribution. As one of the applications of
the present formalism, we theoretically derive the physical temperature
which coincides with that already obtained in accordance with the
generalized zeroth law of thermodynamics.
\end{abstract}

\date{\today }
\startpage{1}
\endpage{1}
\maketitle

\section{Introduction}

Tsallis statistics has been studied as a generalization of the traditional
Boltzmann-Gibbs statistics with huge number of applications since its birth 
\cite{Ts88}\cite{AO01}\cite{GT04}. The most fundamental formalism in Tsallis
statistics has been based on the maximum entropy principle (MEP) for Tsallis
entropy along the Jaynes' original ideas \cite{Ja57a}\cite{Ja57b}\cite{Ja83}%
\cite{TMP98}\cite{MNPP00}. Tsallis entropy is defined by%
\begin{equation}
S_{q}:=-k\sum\limits_{i=1}^{W}p_{i}^{q}\ln _{q}p_{i}\quad \left( q\in 
\mathbb{R}\right) 
\end{equation}%
where $k$ is a positive constant and $\ln _{q}x$ is the $q$-\textit{%
logarithm function}:%
\begin{equation}
\ln _{q}x:=\frac{x^{1-q}-1}{1-q}\quad \left( x>0,q\in \mathbb{R}\right) .
\end{equation}%
Among the MEPs for Tsallis entropy, the following formalism has been
well-known and often applied since 1998 \cite{TMP98}. The obtained
distribution in \cite{TMP98}\ by the MEP under the constraint:%
\begin{eqnarray}
\sum\limits_{i=1}^{W}p_{i} &=&1,  \label{constraint1} \\
\frac{\sum_{i=1}^{W}p_{i}^{q}\varepsilon _{i}}{\sum_{j=1}^{W}p_{j}^{q}}
&=&U_{q}  \label{constraint2}
\end{eqnarray}%
yields the following form: 
\begin{equation}
p_{i}^{\left( \text{TMP}\right) }=\frac{1}{Z_{q}^{\left( \text{TMP}\right)
}\left( \beta \right) }\exp _{q}\left( -\frac{\beta }{\sum_{j=1}^{W}\left(
p_{j}^{\left( \text{TMP}\right) }\right) ^{q}}\left( \varepsilon
_{i}-U_{q}\right) \right)   \label{TMP-distribuion}
\end{equation}%
where $Z_{q}^{\left( \text{TMP}\right) }\left( \beta \right) $ is the
generalized partition function:%
\begin{equation}
Z_{q}^{\left( \text{TMP}\right) }\left( \beta \right)
:=\sum\limits_{i=1}^{W}\exp _{q}\left( -\frac{\beta }{\sum_{j=1}^{W}\left(
p_{j}^{\left( \text{TMP}\right) }\right) ^{q}}\left( \varepsilon
_{i}-U_{q}\right) \right) ,
\end{equation}%
$\beta $ is the Lagrange multiplier associated with the energy constraint (%
\ref{constraint2}) and $\exp _{q}$ is the $q$-\textit{exponential function}
defined by%
\begin{equation}
\exp _{q}\left( x\right) :=\left\{ 
\begin{array}{ll}
\left[ 1+\left( 1-q\right) x\right] ^{\frac{1}{1-q}} & \text{if }1+\left(
1-q\right) x>0, \\ 
0 & \text{otherwise}%
\end{array}%
\right. \quad \left( x\in \mathbb{R},q\in \mathbb{R}\right) 
\end{equation}%
which is the inverse function of the $q$-logarithm function $\ln _{q}x$. In
the above formulation (\ref{TMP-distribuion}), we should pay attention to
the fact that the obtained distribution $p_{i}^{\left( \text{TMP}\right) }$
is also included in the right side of (\ref{TMP-distribuion}) itself. In
other words, the formula (\ref{TMP-distribuion}) is a \textit{%
self-referential} function of $p_{i}^{\left( \text{TMP}\right) }$. From a
mathematical point of view, a form of a self-referential function \textit{%
cannot be uniquely determined} in general. In fact, much simpler
distribution as the solution of the same MEP as above is theoretically
obtained as follows:%
\begin{equation}
p_{i}^{\left( e\right) }=\frac{1}{Z_{q}^{\left( e\right) }\left( \beta
_{q}\right) }\exp _{q}\left( -\beta _{q}\left( \varepsilon _{i}-U_{q}\right)
\right)   \label{Suyari-distribution}
\end{equation}%
where $Z_{q}^{\left( e\right) }\left( \beta _{q}\right) $ is the generalized
partition function:%
\begin{equation}
Z_{q}^{\left( e\right) }\left( \beta _{q}\right) :=\sum\limits_{i=1}^{W}\exp
_{q}\left( -\beta _{q}\left( \varepsilon _{i}-U_{q}\right) \right) ,
\label{Suyari-partition function}
\end{equation}%
$\beta _{q}$ is defined by%
\begin{equation}
\beta _{q}:=\frac{q}{q+\left( 1+\alpha \right) \left( 1-q\right) }\beta ,
\label{beta_q}
\end{equation}%
and $\alpha $ is the Lagrange multiplier associated with the normalization
constraint (\ref{constraint1}) \cite{WS05b}. The index \textquotedblleft $%
\left( e\right) $\textquotedblright\ stands for the equilibrium state in
contrast to \textquotedblleft TMP\textquotedblright\ in order to avoid a
confusion in our discussion. Clearly, the latter distribution (\ref%
{Suyari-distribution}) is \textit{not} a self-referential function of $%
p_{i}^{\left( e\right) }$, which means that this distribution (\ref%
{Suyari-distribution}) is the \textit{unique} solution of the above MEP. In
particular, the distribution (\ref{Suyari-distribution}) is obviously much
simpler and more natural than the usual one (\ref{TMP-distribuion}) from
statistical mechanical point of view. Moreover, $\dfrac{1}{k\beta _{q}}$
theoretically coincides with the \textit{physical temperature} already
obtained in accordance with the generalized zeroth law of thermodynamics 
\cite{AMPP01}\cite{To03}. The physical temperature was first considered in 
\cite{Abe99}\cite{Rama00} and provides us with very significant
interpretation on the observed data in statistical physics such as
astrophysics \cite{Hansen05}, solid-state physics \cite{Hasegawa05} and so on

The purpose of this paper is to show 2 theoretical results: (i) the
derivation of the unique solution (\ref{Suyari-distribution}) from the above
MEP, (ii) the derivation of the physical temperature from our formalism.
After showing these 2 derivations, we discuss the \textit{normalization}
required from the algebra of the $q$-product in Tsallis statistics, which
helps us leading to our present formalism.

\section{Derivation of the unique solution of the MEP in Tsallis statistics}

In this section, we rigorously derive the distribution (\ref%
{Suyari-distribution}) for the MEP as stated in the previous section.

Let $\Phi _{q}$ be defined by a function of $p_{i},\alpha ,\beta $ as
follows: 
\begin{eqnarray}
\Phi _{q}\left( p_{i},\alpha ,\beta \right) &:&=\frac{S_{q}}{k}-\alpha
\left( \sum_{i=1}^{W}p_{i}-1\right) -\beta \frac{\sum_{i=1}^{W}p_{i}^{q}%
\left( \varepsilon _{i}-U_{q}\right) }{\sum_{j=1}^{W}p_{j}^{q}} \\
&=&-\sum\limits_{i=1}^{W}p_{i}^{q}\ln _{q}p_{i}-\alpha \left(
\sum_{i=1}^{W}p_{i}-1\right) -\beta \frac{\sum_{i=1}^{W}p_{i}^{q}\left(
\varepsilon _{i}-U_{q}\right) }{\sum_{j=1}^{W}p_{j}^{q}}.
\end{eqnarray}%
According to the Lagrange multiplier method, the differentials of $\Phi _{q}$
with respect to $p_{i},\alpha ,\beta $ are required to satisfy 
\begin{align}
\frac{\partial \Phi _{q}\left( p_{i},\alpha ,\beta \right) }{\partial p_{i}}%
& =-qp_{i}^{q-1}\ln _{q}p_{i}-1-\alpha -\beta \frac{qp_{i}^{q-1}\left(
\varepsilon _{i}-U_{q}\right) }{\sum_{j=1}^{W}p_{j}^{q}}=0\quad \left(
i=1,\cdots ,W\right) ,  \label{constraint2-1} \\
\frac{\partial \Phi _{q}\left( p_{i},\alpha ,\beta \right) }{\partial \alpha 
}& =1-\sum_{i=1}^{W}p_{i}=0,  \label{constraint2-2} \\
\frac{\partial \Phi _{q}\left( p_{i},\alpha ,\beta \right) }{\partial \beta }%
& =-\frac{\sum_{i=1}^{W}p_{i}^{q}\left( \varepsilon _{i}-U_{q}\right) }{%
\sum_{j=1}^{W}p_{j}^{q}}=0.  \label{constraint2-3}
\end{align}%
The first requirement (\ref{constraint2-1}) is rearranged to the following
equation with respect to $\ln _{q}p_{i}$. 
\begin{equation}
\frac{q+\left( 1+\alpha \right) \left( 1-q\right) }{q}\ln _{q}p_{i}=\frac{%
-\left( 1+\alpha \right) }{q}-\beta \frac{\left( \varepsilon
_{i}-U_{q}\right) }{\sum_{j=1}^{W}p_{j}^{q}}  \label{rearrange_lnqp}
\end{equation}%
Thus, we obtain 
\begin{align}
\ln _{q}p_{i}& =\frac{q}{q+\left( 1+\alpha \right) \left( 1-q\right) }\left( 
\frac{-\left( 1+\alpha \right) }{q}-\beta \frac{\left( \varepsilon
_{i}-U_{q}\right) }{\sum_{j=1}^{W}p_{j}^{q}}\right) \\
& =C_{q}-\beta _{q}\frac{\left( \varepsilon _{i}-U_{q}\right) }{%
\sum_{j=1}^{W}p_{j}^{q}}  \label{ln_pi}
\end{align}%
where $C_{q}$ and $\beta _{q}$ are defined by 
\begin{align}
C_{q}& :=\frac{-\left( 1+\alpha \right) }{q+\left( 1+\alpha \right) \left(
1-q\right) }, \\
\beta _{q}& :=\frac{q}{q+\left( 1+\alpha \right) \left( 1-q\right) }\beta .
\end{align}%
Using $C_{q}$ and $\beta _{q}$, $p_{i}$ in (\ref{ln_pi}) is expanded as
follows: 
\begin{align}
p_{i}& =\exp _{q}\left[ C_{q}-\beta _{q}\frac{\left( \varepsilon
_{i}-U_{q}\right) }{\sum_{j=1}^{W}p_{j}^{q}}\right]  \label{pi_cqbq-1} \\
& =\left( 1+\left( 1-q\right) \left( C_{q}-\beta _{q}\frac{\left(
\varepsilon _{i}-U_{q}\right) }{\sum_{j=1}^{W}p_{j}^{q}}\right) \right) ^{%
\frac{1}{1-q}}  \label{pi_cqbq-2} \\
& =\left( 1+\left( 1-q\right) C_{q}\right) ^{\frac{1}{1-q}}\left( 1+\left(
1-q\right) \left( -\beta _{q}\frac{\left( \varepsilon _{i}-U_{q}\right) }{%
\left( 1+\left( 1-q\right) C_{q}\right) \sum_{j=1}^{W}p_{j}^{q}}\right)
\right) ^{\frac{1}{1-q}}  \label{pi_cqbq-3} \\
& =\exp _{q}\left( C_{q}\right) \exp _{q}\left( -\beta _{q}\frac{\left(
\varepsilon _{i}-U_{q}\right) }{\left( \exp _{q}\left( C_{q}\right) \right)
^{1-q}\sum_{j=1}^{W}p_{j}^{q}}\right)  \label{pi_cqbq-4}
\end{align}%
Substitution of (\ref{pi_cqbq-4}) into the second requirement (\ref%
{constraint2-2}) implies%
\begin{eqnarray}
1 &=&\left( \exp _{q}\left( C_{q}\right) \right)
^{1-q}\sum_{i=1}^{W}p_{i}^{q}\left( 1+\left( 1-q\right) \left( -\beta _{q}%
\frac{\left( \varepsilon _{i}-U_{q}\right) }{\left( 1+\left( 1-q\right)
C_{q}\right) \sum_{j=1}^{W}p_{j}^{q}}\right) \right) \\
&=&\left( \exp _{q}\left( C_{q}\right) \right) ^{1-q}\sum_{i=1}^{W}p_{i}^{q}
\end{eqnarray}%
where the third requirement (\ref{constraint2-3}) is used. Therefore, $p_{i}$
obtained in (\ref{pi_cqbq-4}) is simplified to%
\begin{equation}
p_{i}=\exp _{q}\left( C_{q}\right) \exp _{q}\left( -\beta _{q}\left(
\varepsilon _{i}-U_{q}\right) \right) .  \label{Suyari-distribution0}
\end{equation}%
Here we replace the notation $\exp _{q}\left( C_{q}\right) $ by the familiar
expression $Z_{q}^{\left( e\right) }\left( \beta _{q}\right) $ satisfying 
\begin{equation}
Z_{q}^{\left( e\right) }\left( \beta _{q}\right) =\frac{1}{\exp _{q}\left(
C_{q}\right) }.  \label{Suyari-partition function1}
\end{equation}%
Then we immediately obtain%
\begin{equation}
\sum_{i=1}^{W}\left( p_{i}^{\left( e\right) }\right) ^{q}=\left(
Z_{q}^{\left( e\right) }\left( \beta _{q}\right) \right) ^{1-q}
\label{q-bunpai kansuu}
\end{equation}%
and 
\begin{equation}
p_{i}^{\left( e\right) }=\frac{1}{Z_{q}^{\left( e\right) }\left( \beta
_{q}\right) }\exp _{q}\left( -\beta _{q}\left( \varepsilon _{i}-U_{q}\right)
\right)  \label{Suyari-distribution1}
\end{equation}%
where the notation $p_{i}$ in (\ref{Suyari-distribution0}) is replaced by $%
p_{i}^{\left( e\right) }$ in accordance with the notation of (\ref%
{Suyari-distribution}).

Note that the important step in the present derivation of (\ref%
{Suyari-distribution}) or (\ref{Suyari-distribution1}) is the rearrangement
of (\ref{constraint2-1}) with respect to \textquotedblleft $\ln _{q}p_{i}$%
\textquotedblright , shown in (\ref{rearrange_lnqp}). If (\ref{constraint2-1}%
) is rearranged with respect to \textquotedblleft $p_{i}$\textquotedblright
, the usual distribution (\ref{TMP-distribuion}) is derived. These 2 kinds
of mathematical rearrangements are almost equivalent with each other, but
they result in a big difference in the obtained distributions in the sense
that one is self-referential and the other is not so. The present derivation
is naturally required from the algebra of the $q$-product in Tsallis
statistics, explained in the section IV.

Tsallis entropy for $\left\{ p_{i}^{\left( e\right) }\right\} $ is
immediately derived from (\ref{q-bunpai kansuu}) as follows:%
\begin{equation}
S_{q}\left( \left\{ p_{i}^{\left( e\right) }\right\} \right) =k\ln
_{q}Z_{q}^{\left( e\right) }\left( \beta _{q}\right) .  \label{entroy for eq}
\end{equation}%
On the other hand, Tsallis entropy $S_{q}\left( \left\{ p_{i}^{\left( \text{%
TMP}\right) }\right\} \right) $ for $\left\{ p_{i}^{\left( \text{TMP}\right)
}\right\} $ has the same form as above, shown in \cite{TMP98}.

\section{Derivation of the physical temperature from our formalism}

\bigskip We derive the physical temperature from our formalism presented in
the previous section. For $\left\{ p_{i}^{\left( e\right) }\right\} $ given
in (\ref{Suyari-distribution}) or (\ref{Suyari-distribution1}), $\dfrac{%
\partial S_{q}\left( \left\{ p_{i}^{\left( e\right) }\right\} \right) }{%
\partial U_{q}}$ is computed as follows: 
\begin{align}
\frac{\partial S_{q}\left( \left\{ p_{i}^{\left( e\right) }\right\} \right) 
}{\partial U_{q}}& =k\frac{\partial }{\partial U_{q}}\ln _{q}Z_{q}^{\left(
e\right) }\left( \beta _{q}\right) \\
& =k\frac{1}{\left( Z_{q}^{\left( e\right) }\left( \beta _{q}\right) \right)
^{q}}\sum_{i=1}^{W}\frac{\partial }{\partial U_{q}}\left[ 1+\left(
1-q\right) \left( -\beta _{q}\left( \varepsilon _{i}-U_{q}\right) \right) %
\right] ^{\frac{1}{1-q}} \\
& =k\beta _{q}\sum_{i=1}^{W}\left( p_{i}^{\left( e\right) }\right) ^{q} \\
& =k\beta _{q}\left( Z_{q}^{\left( e\right) }\left( \beta _{q}\right)
\right) ^{1-q} \\
& =k\beta _{q}\left( 1+\frac{1-q}{k}S_{q}\left( \left\{ p_{i}^{\left(
e\right) }\right\} \right) \right)
\end{align}%
Therefore, if $\beta _{q}$ is denoted by means of a new parameter $T_{q}$
such as 
\begin{equation}
\beta _{q}=\frac{1}{kT_{q}},
\end{equation}%
then%
\begin{equation}
T_{q}=\left( 1+\frac{1-q}{k}S_{q}\left( \left\{ p_{i}^{\left( e\right)
}\right\} \right) \right) \left( \frac{\partial S_{q}\left( \left\{
p_{i}^{\left( e\right) }\right\} \right) }{\partial U_{q}}\right) ^{-1}.
\label{suyari_temperature}
\end{equation}

Note that $T_{q}$ is self-consistently derived from our formalism only.

The obtained $T_{q}$ above coincides with the physical temperature $T_{\text{%
phys}}$ derived by Abe at al \cite{AMPP01} in the following sense. In \cite%
{AMPP01}, using the pseudoadditivity of Tsallis entropy and fixed internal
energy constraint only, the physical temperature $T_{\text{phys}}$ is
derived in accordance with the zeroth law of thermodynamics. More precisely,
the pseudoadditivity of Tsallis entropy for thermal equilibrium requires 
\begin{eqnarray}
0 &=&\delta S_{q}\left( A,B\right)  \\
&=&\left( 1+\frac{1-q}{k}S_{q}\left( B\right) \right) \frac{\partial
S_{q}\left( A\right) }{\partial U_{q}\left( A\right) }\delta U_{q}\left(
A\right) +\left( 1+\frac{1-q}{k}S_{q}\left( A\right) \right) \frac{\partial
S_{q}\left( B\right) }{\partial U_{q}\left( B\right) }\delta U_{q}\left(
B\right)   \label{Abe1}
\end{eqnarray}%
where $A$ and $B$ are two independent subsystems composing the total system.
On the other hand, the fixed internal energy constraint requires%
\begin{equation}
0=\delta U_{q}\left( A,B\right) =\delta U_{q}\left( A\right) +\delta
U_{q}\left( B\right) .  \label{Abe2}
\end{equation}%
These requirements (\ref{Abe1}) and (\ref{Abe2}) yields 
\begin{equation}
\left( 1+\frac{1-q}{k}S_{q}\left( B\right) \right) \left( \frac{\partial
S_{q}\left( B\right) }{\partial U_{q}\left( B\right) }\right) ^{-1}=\left( 1+%
\frac{1-q}{k}S_{q}\left( A\right) \right) \left( \frac{\partial S_{q}\left(
A\right) }{\partial U_{q}\left( A\right) }\right) ^{-1}.
\end{equation}%
Therefore, the physical temperature $T_{\text{phys}}$ is defined in \cite%
{AMPP01} as follows: 
\begin{equation}
T_{\text{phys}}:=\left( 1+\frac{1-q}{k}S_{q}\right) \left( \frac{\partial
S_{q}}{\partial U_{q}}\right) ^{-1}  \label{Abe's temperature}
\end{equation}%
This coincides with $T_{q}$ in (\ref{suyari_temperature}) which is obtained
in our formalism only. In other words, $T_{q}$ and $T_{\text{phys}}$ are
independently derived in each formalism and coincide with each other.

Note that in the formula (5) of \cite{AMPP01} the Lagrange multiplier $\beta 
$ is used for the definition of the physical temperature $T_{\text{phys}}$,
but actually (\ref{Abe's temperature}) is the most general formula of the
physical temperature $T_{\text{phys}}$ derived in \cite{AMPP01}. See also
(1) in \cite{To03}.

\section{Formalism required from the algebra of the $q$-product in Tsallis
statistics}

\bigskip In our formalism, we applied the mathematical structure of the $q$%
-product in Tsallis statistics to the present derivation of (\ref%
{Suyari-distribution}) or (\ref{Suyari-distribution1}). This section
describes the useful technique and insight for finding formulae in Tsallis
statistics through concrete examples.

The $q$-product \cite{NMW03}\cite{Bo04} is defined for positive numbers $%
x,\,y\in \mathbb{R}^{+}$ as follows:%
\begin{equation}
x\times _{q}y:=\left\{ 
\begin{array}{ll}
\left[ x^{1-q}+y^{1-q}-1\right] ^{\frac{1}{1-q}}, & \text{if }%
x>0,\,y>0,\,x^{1-q}+y^{1-q}-1>0, \\ 
0, & \text{otherwise.}%
\end{array}%
\right.   \label{def of q-product}
\end{equation}%
The definition of the $q$-product originates from the requirement of the
following satisfactions:%
\begin{align}
\ln _{q}\left( x\times _{q}y\right) & =\ln _{q}x+\ln _{q}y,
\label{property of ln_q} \\
\exp _{q}\left( x\right) \times _{q}\exp _{q}\left( y\right) & =\exp
_{q}\left( x+y\right) .  \label{property of exp_q}
\end{align}%
As the inverse function of the $q$-product, $q$-ratio \cite{NMW03}\cite{Bo04}
is introduced:%
\begin{equation}
x\diagup _{q}y:=\left\{ 
\begin{array}{lll}
\left[ x^{1-q}-y^{1-q}+1\right] ^{\frac{1}{1-q}}, &  & \text{if }%
x>0,\,y>0,\,x^{1-q}-y^{1-q}+1>0, \\ 
0, &  & \text{otherwise}%
\end{array}%
\right. 
\end{equation}%
to satisfy the following requirements:%
\begin{align}
\ln _{q}x\diagup _{q}y& =\ln _{q}x-\ln _{q}y, \\
\exp _{q}\left( x\right) \diagup _{q}\exp _{q}\left( y\right) & =\exp
_{q}\left( x-y\right) .
\end{align}%
The $q$-product plays a very important role in the formalism in Tsallis
statistics such as law of error \cite{ST0312}, $q$-Stirling's formula \cite%
{Su04a}, $q$-multinomial coefficient \cite{Su04b}, Pascal triangle in
Tsallis statistics \cite{Su04b} and so on. Not only in these results but
also in most of subjects in Tsallis statistics the $q$-product is often
appeared. The mathematical insight into the algebra of the $q$-product helps
us leading to the results we want. In particular, in the formulation by
means of the $q$-product, \textit{normalization} should be taken very
carefully. Such examples are shown below.

\begin{example}
As the simplest example of such situations, we solve the following
differential equation:%
\begin{equation}
\frac{dy}{dx}=y^{q}
\end{equation}%
where Tsallis often took the above equation as one of the introductory
characterizations of the fundamental formulae in Tsallis statistics \cite%
{GT04}. The above differential equation is solved as follows:%
\begin{equation}
\ln _{q}y=x+C,\quad \text{that is,\quad }y=\exp _{q}\left( x+C\right) 
\label{expansion_q-product-1}
\end{equation}%
where $C$ is a constant. Here we should pay attention to the argument $x+C$
in $\exp _{q}$. A constant $C$ is not set in front of $\exp _{q}$, but in
the argument of $\exp _{q}$. This solution $y$ can be rewritten by means of
the $q$-product: 
\begin{equation}
y=\exp _{q}\left( x\right) \times _{q}\exp _{q}\left( C\right) .
\label{expansion_q-product}
\end{equation}%
Thus, $\exp _{q}\left( C\right) $ plays a role of \textit{normalization} in
the sense of the $q$-product. If we need to normalize it with respect to $x$
in the sense of the usual product, the above solution $y$ is expanded as
follows:%
\begin{eqnarray}
y &=&\left[ 1+\left( 1-q\right) \left( x+C\right) \right] ^{\frac{1}{1-q}} \\
&=&\exp _{q}\left( C\right) \exp _{q}\left( \frac{x}{1+\left( 1-q\right) C}%
\right) .  \label{expansion_product}
\end{eqnarray}%
Note that $x$ and $C$ in (\ref{expansion_q-product}) are factorized using
the $q$-product, but $x$ and $C$ in (\ref{expansion_product}) for the same $y
$ as (\ref{expansion_q-product}) cannot be factorized using the usual
product in general. This transformation between (\ref{expansion_q-product})
and (\ref{expansion_product}) is very useful in some formulations in Tsallis
statistics (e.g., (\ref{pi_cqbq-1})-(\ref{pi_cqbq-4}) in this manuscript or
(46)-(49) in \cite{ST0312}).
\end{example}

\begin{example}
Keeping the fundamental expansion (\ref{expansion_q-product-1}) in mind, we
solve the same MEP for Shannon entropy. The requirement (\ref{constraint2-1}%
) when $q=1$:%
\begin{equation}
\frac{\partial \Phi _{1}\left( p_{i},\alpha ,\beta \right) }{\partial p_{i}}%
=-\ln p_{i}-1-\alpha -\beta \left( \varepsilon _{i}-U_{q}\right) =0
\label{constraint_Shannon}
\end{equation}%
implies 
\begin{eqnarray}
\ln p_{i} &=&-1-\alpha -\beta \left( \varepsilon _{i}-U_{q}\right)
\label{expansion_Shannon-1} \\
\text{that is,\quad }p_{i} &=&\exp \left( -\beta \left( \varepsilon
_{i}-U_{q}\right) -1-\alpha \right)  \label{expansion_Shannon} \\
&=&\exp \left( -\beta \left( \varepsilon _{i}-U_{q}\right) \right) \times
\exp \left( -1-\alpha \right) .
\end{eqnarray}
\end{example}

These 2 examples tell us that the representation%
\begin{equation}
y=\exp _{q}\left( x+C\right) 
\end{equation}%
such as (\ref{expansion_q-product-1}) and (\ref{expansion_Shannon}) is the
most \textit{fundamental} form in Tsallis statistics (See also (46) in \cite%
{ST0312}). Therefore, in order to obtain a fundamental form such as (\ref%
{expansion_q-product-1}) in each problem in Tsallis statistics, we should
derive an equation with respect to \textquotedblleft ln$_{q}p_{i}$%
\textquotedblright , not \textquotedblleft $p_{i}$\textquotedblright . In
fact, when $q=1$, ln$p_{i}$ is derived from the requirement (\ref%
{constraint_Shannon}) at first. When $q=1$, there is no difference between
these 2 kinds of reformulations by means of \textquotedblleft ln$_{q}p_{i}$%
\textquotedblright\ and \textquotedblleft $p_{i}$\textquotedblright .
However, when $q\neq 1,$ we can see the difference between these 2
formulations in the obtained distributions: $p_{i}^{\left( e\right) }$ in (%
\ref{Suyari-distribution}) and $p_{i}^{\left( \text{TMP}\right) }$ in (\ref%
{TMP-distribuion}). The former is derived from the equation (\ref%
{rearrange_lnqp}) of ln$_{q}p_{i}$, but the latter is directly derived from
the equation (\ref{constraint2-1}) of $p_{i}$. This difference comes from
the distinction between the algebra of the $q$-product in Tsallis statistics
and that of the usual product in Boltzmann-Gibbs statistics. In some sense,
this distinction is purely mathematical, but it results in a serious
influence on the formalism in statistical physics.

\section{Conclusion}

We derive the unique non self-referential $q$-canonical distribution (\ref%
{Suyari-distribution}) from the maximum entropy principle (MEP) in Tsallis
statistics by taking account of the algebra required from the $q$-product.
The obtained distribution $p_{i}^{\left( e\right) }$ in (\ref%
{Suyari-distribution}) is obviously much simpler and more natural than the
usual $p_{i}^{\left( \text{TMP}\right) }$ in (\ref{TMP-distribuion}).
Moreover, we derive the physical temperature from our formalism, which
coincides with that already obtained in another way by Abe et al in
accordance with the generalized zeroth law of thermodynamics.

Recently, we have presented the following fundamental results in Tsallis
statistics:

\begin{enumerate}
\item Axioms and the uniqueness theorem for the nonextensive entropy \cite%
{Su04c}

\item Law of error in Tsallis statistics \cite{ST0312}

\item $q$-Stirling's formula in Tsallis statistics \cite{Su04a}

\item $q$-multinomial coefficient in Tsallis statistics \cite{Su04b}

\item Central limit theorem in Tsallis statistics (numerical evidence only) 
\cite{Su04b}

\item Pascal triangle in Tsallis statistics \cite{Su04b}

\item Maximum entropy principle in Tsallis statistics [the present paper \& 
\cite{WS05b}]
\end{enumerate}

Most of the above results are derived from the algebra of the $q$-product.
This means that $q$-product is indispensable to the formalism in Tsallis
statistics. On the other hand, Tsallis statistics has been found to be the
well-organized statistical mechanics with many fruitful applications as a
nice generalization of the traditional Boltzmann-Gibbs statistics. In our
near future, we strongly believe these formalism in Tsallis statistics
brings about significant influences on other sciences such as information
theory and network theory.

\bigskip

\textbf{Note added }After uploading the 1st version of the preprint
(cond-mat/0502298), we (Dr. Wada and H.S.) exchanged our results (the 1st
version of this preprint and \cite{WS05b}) and confimed that we
independently obtained the same $q$-canonical distribution (\ref%
{Suyari-distribution}) in a different way.

\bigskip

\textbf{Acknowledgment} We gratefully acknowledge Dr. Tatsuaki Wada for some
informations on his result.


\begin{thebibliography}{99}
\bibitem{Ts88} C. Tsallis, Possible generalization of Boltzmann-Gibbs
statistics, J. Stat. Phys. 52, 479-487 (1988).

\bibitem{AO01} S. Abe and Y. Okamoto, eds., Nonextensive Statistical
Mechanics and Its Applications (Springer-Verlag, Heidelberg, 2001).

\bibitem{GT04} M. Gell-Mann and C. Tsallis, eds., Nonextensive Entropy:
Interdisciplinary Applications (Oxford Univ. Press, New York, 2004).

\bibitem{Ja57a} E. T. Jaynes, Information theory and statistical mechanics,
Phys. Rev. 106, 620-630 (1957).

\bibitem{Ja57b} E. T. Jaynes, Information theory and statistical mechanics
II, Phys. Rev. 108, 171-190 (1957).

\bibitem{Ja83} E.T. Jaynes, Papers on Probability, Statistics and
Statistical Physics, edited by R.D. Rosenkrantz (D. Reidel, Boston, 1983).

\bibitem{TMP98} C. Tsallis, R.S. Mendes, A.R. Plastino, The role of
constraints within generalized nonextensive statistics, Physica A 261,
534-554 (1998).

\bibitem{MNPP00} S. Mart\'{i}nez, F. Nicol\'{a}s, F. Pennini and A.
Plastino, Tsallis' entropy maximization procedure revisited, Physica A 286,
489-502 (2000).

\bibitem{WS05b} T. Wada, Dr. A.M. Scarfone, A non self-referential
expression of Tsallis' probability distribution function, LANL e-print
cond-mat/0502394.

\bibitem{AMPP01} S. Abe, S. Mart\'{\i}nez, F. Pennini, A. Plastino,
Nonextensive thermodynamic relations, Phys. Lett. A 281, 126-130 (2001).

\bibitem{To03} R. Toral, On the definition of physical temperature and
pressure for nonextensive thermostatistics, Physica A 317, 209-212 (2003).

\bibitem{Abe99} S. Abe, Correlation induced by Tsallis' nonextensivity,
Physica A 269, 403-409 (1999).

\bibitem{Rama00} S. K. Rama, Tsallis statistics: averages and a physical
interpretation of the Lagrange multiplier $\beta ,$ Phys. Lett. A 276,
103--108 (2000).

\bibitem{Hansen05} S. H. Hansen, Cluster temperatures and non-extensive
thermo-statistics, New Astronomy (in press), LANL e-print astro-ph/0501393.

\bibitem{Hasegawa05} H. Hasegawa, Nonextensive thermodynamics of a cluster
consisting of $M$ Hubbard dimers ($M=1,2,3$ and $\infty $), LANL e-print
cond-mat/0501126.

\bibitem{NMW03} L. Nivanen, A. Le M\'{e}haut\'{e}, Q.A. Wang, Generalized
algebra within a nonextensive statistics, Rep. Math. Phys. 52, 437-444
(2003).

\bibitem{Bo04} E.P. Borges, A possible deformed algebra and calculus
inspired in nonextensive thermostatistics, Physica A 340, 95--101 (2004).

\bibitem{ST0312} H. Suyari, M. Tsukada, Law of error in Tsallis statistics,
IEEE Trans. Inform. Theory., vol.51, pp.753-757 (2005).

\bibitem{Su04a} H. Suyari, $q$-Stirling's formula in Tsallis statistics,
LANL e-print cond-mat/0401541.

\bibitem{Su04b} H. Suyari, Mathematical structure derived from the $q$%
-multinomial coefficient in Tsallis statistics, LANL e-print
cond-mat/0401546.

\bibitem{Su04c} H. Suyari, Generalization of Shannon-Khinchin axioms to
nonextensive systems and the uniqueness theorem for the nonextensive
entropy, IEEE Trans. Inform. Theory., vol.50, pp.1783-1787 (2004).
\end{thebibliography}
\end{document}